\documentclass[aps,prb,amsfonts,amssymb,twocolumn,amsmath,preprintnumbers,nofootinbib,floatfix,superscriptaddress,
showpacs]{revtex4-1}
\usepackage[dvips]{graphics}
\usepackage{graphicx}
\usepackage{bm}

\begin{document}

\title{In-plane FFLO instability in superconductor-normal metal bilayer system under non-equilibrium quasiparticle distribution}
\author{I. V. Bobkova}
\affiliation{Institute of Solid State Physics, Chernogolovka,
Moscow reg., 142432 Russia}
\affiliation{Moscow Institute of Physics and Technology, Dolgoprudny, 141700 Russia}
\author{A. M. Bobkov}
\affiliation{Institute of Solid State Physics, Chernogolovka,
Moscow reg., 142432 Russia}

\date{\today}

\begin{abstract}
It is predicted that a new class of systems - superconductor/normal metal (S/N) heterostructures can reveal the in-plane Fulde-Ferrel-Larkin-Ovchinnikov (FFLO) instability under nonequilibrium conditions at temperatures close to the critical temperature. It does not require any Zeeman interaction in the system. For S/N heterostructures under non-equilibrium distribution there is a natural easily adjustable  parameter -  voltage, which can control the FFLO-state. This FFLO-state can be of different types: plane wave, stationary wave and, even, 2D-structures are possible. Some types of the FFLO-state are accompanied by the magnetic flux, which can be observed experimentally. All the types of the FFLO-state can be revealed through the temperature dependence of the linear response of the system on the applied magnetic field near $T_c$, which strongly differs from that one for the homogeneous state.  
\end{abstract}

\maketitle

There are two mechanisms of superconductivity
destruction by a magnetic field: orbital effect and the Zeeman interaction of electron spins with
the magnetic field. Usually the orbital effect is more restrictive. However there are several classes of systems,
where the orbital effect is strongly weakened (systems with large
effective mass of electrons \cite{bianchi03,capan04}, thin films and layered superconductors
under in-plane magnetic field \cite{uji01}) or even completely absent (superconductor/ferromagnet (S/F) heterostructures \cite{buzdin05,bergeret05}). Then the Zeeman interactions of electron spins with a magnetic or an exchange field is responsible for the superconductivity 
destruction. 

The behavior of a superconductor
with a homogeneous exchange field $h$ was studied long ago \cite{larkin64,fulde64,sarma63,maki68}. It
was found that homogeneous superconducting state becomes
energetically unfavorable above the paramagnetic
(Pauli) limit $h=\Delta_0/\sqrt 2$, where $\Delta_0$ is the zero-temperature superconducting gap. As it was predicted by 
Larkin and Ovchinnikov \cite{larkin64} and by Fulde and Ferrell \cite{fulde64}, in a
narrow region of exchange fields exceeding this value superconductivity
can appear as an inhomogeneous state
with a spatially modulated Cooper pair wave function
(FFLO-state). 

Now there is a growing body of experimental evidence for the FFLO phase, generated by the applied magnetic field, 
reported from various measurements \cite{singleton00,tanatar02,bianchi02,miclea06,uji06,shinagawa07,lortz07,cho09,wright11,bergk11,coniglio11,tarantini11,gebre11,agosta12,uji12}. However, any unambiguous experimental results, which can be interpreted only as a fingerprint of the FFLO-state, 
are not reported by now.

On the other hand, it has been predicted recently \cite{mironov12} that the FFLO-state can be realized in S/F heterostructures, where S is a singlet s-wave superconductor. Here we mean the so-called in-plane FFLO-state, where the superconducting order parameter profile is modulated along the layers. It should be distinguished from
the normal to the S/F interface oscillations of the condensate wave function in the ferromagnetic layer, which are well investigated as theoretically, so as experimentally \cite{buzdin05,bergeret05,sidorenko09}. 

In this paper we show that the in-plane FFLO-state can be the most energetically favorable state in S/N heterostructures under the non-equilibrium quasiparticle distribution and propose a way to observe it. The exchange field is absent in S/N heterostructures.  Correspondingly, there is no Zeeman interaction   without applied magnetic field. The transition to the FFLO-state occurs due to creation of a double-step electron distribution in the bilayer. This non-equilibrium state can be reached by changing the chemical potentials of additional electrodes in opposite directions by applying a control voltage \cite{pothier97,baselmans99}. To the best of our knowledge, there are a very few proposals of the FFLO-state in non-magnetic systems (for example, a current-driven FFLO-state in 2D superconductors with Fermi surface nesting \cite{doh06}, in unconventional superconducting films \cite{vorontsov09} and in nonequilibrium N/S/N heterostructures at low enough temperatures \cite{volkov09}). The effect considered here strongly differs from the one discussed in Ref.~\onlinecite{volkov09}. It was demonstrated in Ref.~\onlinecite{volkov09} that a superconductor under the particular quasiparticle distribution is very similar to the superconductor in the uniform exchange field. Therefore, the LOFF-state can be realized in this system. It is only possible at low temperatures, as it is known for superconductors in the uniform exchange field \cite{saint-james69}. Such a system is not enough to obtain the FFLO-state at temperatures close to $T_c$. Here we show that two essential components: non-equilibrium quasiparticle distribution and the proximity between a superconducting film and a normal film of the particular finite width allow us to obtain the FFLO-state near $T_c$. The possibility to obtain the FFLO-state at temperatures close to $T_c$ is of great interest at least for two reasons: (i) we propose a way to reveal this FFLO-state through the temperature dependence of its linear response on the applied magnetic field near $T_c$; (ii) the orbital effect of the applied magnetic field is highly non-trivial in the FFLO-state: it can enhance $T_c$ instead of its suppression \cite{bobkov13}. 

In addition we propose an alternative way to generate the FFLO-state in S/N heterostructures. It can occur due to creation of two shifted Fermi-surfaces for spin-up and spin-down electrons if the spin imbalance is generated in the system. 

\begin{figure}[!tbh]
  \centerline{\includegraphics[clip=true,width=1.2in]{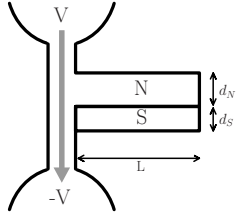}}
    \caption{Sketch of the system under consideration.}   
\label{sketch}
\end{figure}  

Now we proceed with the microscopic calculations of the FFLO critical temperature of the S/N bilayer under non-equilibrium conditions. The sketch of the system is shown in Fig.~\ref{sketch}. As we consider a non-equilibrium system, we make use of Keldysh framework of the quasiclassical theory. In our calculations we assume that (i) S is a singlet s-wave superconductor; (ii) the system is in the dirty limit, so the quasiclassical Green's function obeys Usadel equations \cite{usadel}; (iii) the thickness $d_S$ of the S layer is less than the superconducting coherence length $\xi_S=\sqrt{D_S/\Delta_0}$. This condition allows us to neglect the variations of the superconducting order parameter and the Green's functions across the S layer; (iv) we work in the vicinity of the critical temperature, so the Usadel equations can be linearized with respect to the anomalous Green's function:

\begin{equation}
D \nabla^2 \hat f^R + 2 i \varepsilon \hat f^R +2 \pi \hat \Delta = 0
\enspace .
\label{usadel} 
\end{equation}
Here $\hat f^R \equiv \hat f^R(\varepsilon,\bm r)$ is the retarded anomalous Green's function. It depends on the quasiparticle energy $\varepsilon$ and the coordinate vector $\bm r=(x, \bm r_\parallel)$, where $x$ is the coordinate normal to the S/N interface and $\bm r_\parallel$ is parallel to the interface [$(yz)$ plane]. $\hat ~$ means that the anomalous Green's function is a $2\times 2$ matrix in spin space. However, here we consider the S/N system without the Zeeman interaction, so the retarded and advanced components of the Green's function have the standard  spin-singlet structure $\hat f^{R,A}=f^{R,A}i\sigma_2$, where $\sigma_2$ is the corresponding Pauli matrix. While we only consider the singlet pairing channel, the same is valid for the superconducting order parameter $\hat \Delta=\Delta i \sigma_2$. The spin structure can only appear in the distribuion function, as it is described below. $D=D_{S(N)}$ stands for the diffusion constant in the superconductor (normal metal). 

Eq.~\ref{usadel} should be supplied by the Kupriyanov-Lukichev boundary conditions \cite{kupriyanov88} at the S/N interface ($x=0$)
\begin{equation}
 \sigma_S \partial_x f^R_S = \sigma_N \partial_x f^R_N = 
g_{NS}\left.(f^R_S-f^R_N)\right|_{x=0},
\label{interface_cond}
\end{equation}
where $\sigma_{S(N)}$ stands for a conductivity of the S(N) layer and $g_{NS}$ is the conductance of the S/N interface. The boundary conditions at the ends of the bilayer are $\left. \partial_x f^R_S \right|_{x=d_S} = \left. \partial_x f^R_N \right|_{x=-d_N}=0 $.

In the FFLO-state the superconducting order parameter and the anomalous Green's function are spatially modulated. We assume that
$\Delta(\bm r)=\Delta\exp(i\bm k\bm r_\parallel)$ and $f(\bm r)=f(x)\exp(i\bm k \bm r_\parallel)$. It is worth to note here that this plane wave is not the only possible type of the spatially modulated FFLO-state, which is allowed in the system. There can be also stationary wave states modulated as $\cos (\bm k\bm r_\parallel)$ and also 2D modulated structures. However, it can be shown that the critical temperature of all these states is the same and only depends on the absolute value of the modulating vector $\bm k$. Further choice of the most energetically favorable configuration is determined by the non-linear terms in the Usadel equation, which are neglected now. So, while we are only interested in the instability point and the critical temperature of the corresponding FFLO-state, we can consider the most simple type of the modulation.

Substituting the modulated Green's function into the Usadel equation we obtain the anomalous Green's functions in the S and N layers:
\begin{equation}
f_S=\frac{i\pi\Delta}{\varepsilon + iD_S k^2/2 + \frac{ig_{NS}D_S\lambda \tanh [\lambda d_N]}{2\sigma_S d_s(\lambda \tanh [\lambda d_N]+g_{NS}/\sigma_N)}}
\enspace ,
\label{fS}
\end{equation}
\begin{equation}
f_N(x)=\frac{(g_{NS}/\sigma_N)\cosh[\lambda(x+d_N)]}{\lambda \sinh [\lambda d_N]+(g_{NS}/\sigma_N)\cosh[\lambda d_N]}f_S
\enspace ,
\label{fN}
\end{equation}
where $\lambda^2=k^2-2i\varepsilon/D_N $. 

The critical temperature of the bilayer should be determined from the self-consistency equation
\begin{equation}
\Delta=\int \limits_{-\omega_D}^{\omega_D} \frac{d\varepsilon}{4\pi} \Lambda {\rm Im} \left[ f^R_S \right] (\varphi_\uparrow+\varphi_\downarrow)
\enspace ,
\label{Tc}
\end{equation}
where $\omega_D$ is the cutoff energy, $\Lambda$ is the dimensionless coupling constant and $\varphi_{\uparrow,\downarrow}$ is the distribution function for spin-up (down) quasiparticles. In order to generate the FFLO-state we need 
\begin{equation}
\varphi_\uparrow + \varphi_\downarrow = \tanh \frac{\varepsilon-eV}{2T}+\tanh \frac{\varepsilon+eV}{2T}
\enspace .
\label{distribution}
\end{equation}
This quasiparticle distribution can be reached in the bilayer in two different ways. (i) If the bilayer is attached to two additional electrodes with a voltage applied between them. We assume that the bilayer length $L$ is shorter than the energy relaxation length. Then the energy distribution of the electrons in the bilayer is given by the superposition of the Fermi-Dirac distributions of the reservoirs \cite{pothier97,baselmans99} and $\varphi_\uparrow = \varphi_\downarrow$. (ii) If an electric current is injected into the bilayer through a ferromagnet, the spin imbalance is generated at the interface between the ferromagnet and the non-magnetic region. This is the so-called Aronov gap \cite{aronov76,tsoi}. It provides the conversion (by spin relaxation processes) of the spin-polarized current, injected from the ferromagnet, into the non spin-polarized current, which can only flow through non ferromagnetic material. The value of the Aronov gap at the interface with the ferromagnet can be estimated as $eV \sim e P j_{inj} \rho l_s$, where $P$ is the degree of spin polarization in the ferromagnet, $j_{inj}$ is the density of the current, injected from the ferromagnet, $\rho$ is the resistivity of the normal metal and $l_s$ is the spin relaxation length in it. The spin relaxation length is usually large in normal metals, so we can assume that our bilayer is shorter than $l_s$ and, consequently, the spin imbalance is spatially constant in it. In this case $\varphi_{\uparrow(\downarrow)}=\tanh[(\varepsilon \mp eV)/2T]$.

\begin{figure}[!tbh]
   \begin{minipage}[b]{0.5\linewidth}
     \centerline{\includegraphics[clip=true,width=1.64in]{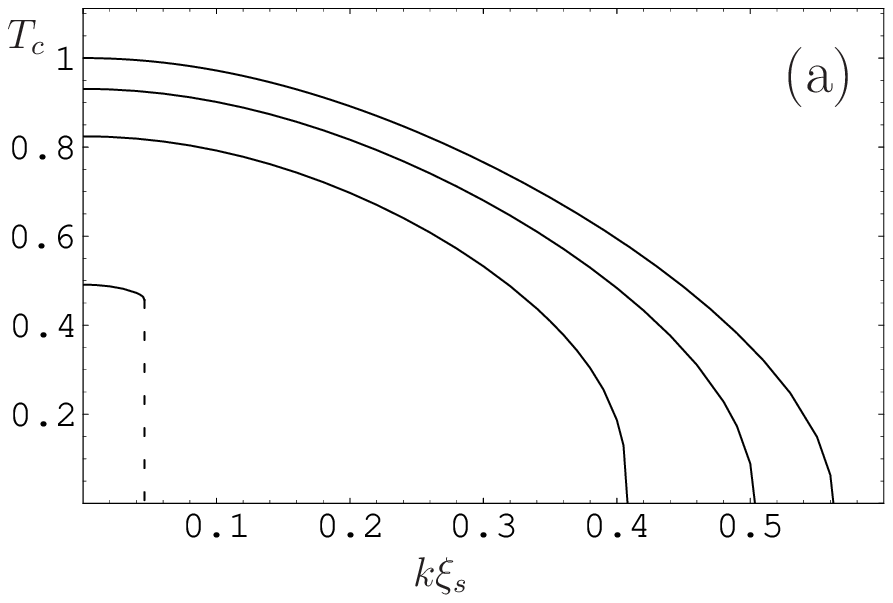}}
     \end{minipage}\hfill
    \begin{minipage}[b]{0.5\linewidth}
   \centerline{\includegraphics[clip=true,width=1.64in]{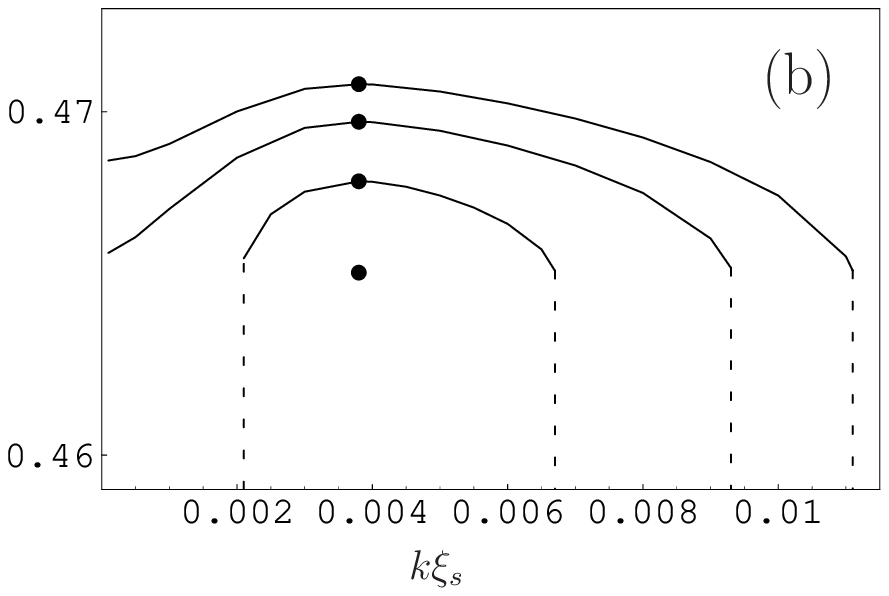}}
  \end{minipage}
   \caption{(a) Dependence of the S/N bilayer critical temperature vs the modulation vector $k$ for different values of $eV$. Points mark $k=k_{opt}$. In units of $T_c(eV=0,k=0)$ $eV$=0; 0.556; 0.833; 1.111 from top to bottom.(b) Enlarged region of panel (a), corresponding to small $k$. $eV$=1.112401; 1.112421; 1.112443; 1.112457 from top to bottom. The other parameters are $\sigma_S/\sigma_N=0.5$, $\xi_Sg_{NS}/\sigma_S=1.0$, $D_S/D_N=0.04$, $d_N=6.0\xi_S$, $d_S=0.8\xi_S$ \cite{note1}.}   
\label{Tc_k}
\end{figure}

The critical temperature of the S/N bilayer as a function of the modulation vector $k$ is represented in Fig.~\ref{Tc_k}. Different curves correspond to different values of the applied voltage $eV$. The curves of most physical interest are in region of small $k$ and narrow interval of $eV$ close to $eV_c$ [See Fig.~\ref{Tc_k}(b)]. The critical voltage $eV_c$ corresponds to the complete destruction of homogeneous superconductivity in our bilayer. It is seen from Fig.~\ref{Tc_k}(b) that if $eV$ is close enough to $eV_c$, the critical temperature of the FFLO-state is higher than $T_c$ of the homogeneous state. That is, the FFLO-state is energetically more favorable. The optimal values of the modulation vector $k_{opt}$, corresponding to the maximal $T_c$, are marked by points. 

\begin{figure}[!tbh]
  \centerline{\includegraphics[clip=true,width=2.0in]{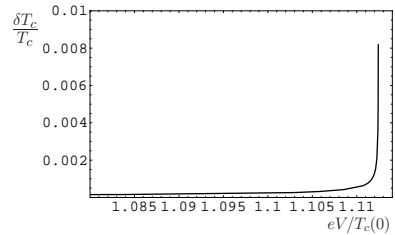}}
   \caption{Dependence of $\delta T_c/T_c=[T_c(k_{opt})-T_c(k=0)]/T_c$ on $eV$. The parameters of the system are the same as in Fig.~\ref{Tc_k}.}   
\label{delta_Tc}
\end{figure}

More detailed analysis shows that for the system under consideration the mean-field $T_c$ is higher for a finite $k$ than for $k=0$ at any $eV$.  Does this mean that the S/N bilayer should be in the FFLO-state even in equilibrium (at $eV=0$)? In order to analyze this question we plot in Fig.~\ref{delta_Tc} the difference between the critical temperature of the FFLO-state corresponding to $k_{opt}$ and the critical temperature of the homogeneous state $\delta T_c/T_c=[T_c(k_{opt})-T_c(k=0)]/T_c$ vs $eV$. As it is seen from Fig.~\ref{delta_Tc}, $\delta T_c/T_c$ is very small for a wide range of $eV$ and only grows sharply in the narrow region near $eV_c$. We have estimated that for small enough voltage biases $\delta T_c/T_c$ does not exceed considerably the Ginzburg number ${\rm Gi}_{\rm 2D}\sim 0.1 /(k_F^2 l d) \approx 10^{-4}\div 10^{-3}$. So, we cannot conclude on the basis of our mean field analysis if the FFLO-state or the homogeneous state is more energetically favorable in this voltage range. However, in the narrow region of $eV$ near $eV_c$ (estimated width $\sim 0.1 \div 1 \mu V$) $\delta T_c/T_c$ exceeds ${\rm Gi}_{\rm 2D}$ at least by the order of magnitude. So, for this voltage region the FFLO-state is indeed more  favorable.  

In addition, there is a narrow voltage region $eV>eV_c$, where homogeneous superconductivity is completely destroyed, but the FFLO-state survives [See Fig.~\ref{Tc_k}(b), where the bottom curve corresponds to $eV>eV_c$].

It is worth noting here that in order to observe the FFLO-state the number of inelastic scatterers should be very small in the system. They can be described by adding the imaginary part $\Gamma$ to the quasiparticle energy $\varepsilon \to \varepsilon+i\Gamma$. Then the condition $\Gamma<D_Sk^2_{opt}/2$ should be fulfilled.    

\begin{figure}[!tbh]
  \centerline{\includegraphics[clip=true,width=2.5in]{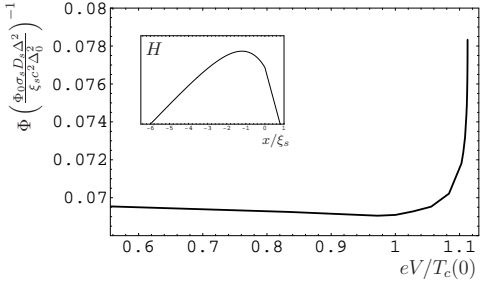}}
   \caption{Dependence of the magnetic flux per unit length along $\bm k$, generated in the plane wave FFLO-state of the S/N bilayer, vs $eV$. Insert: the spatial profile of the corresponding magnetic filed.}   
\label{flux}
\end{figure}

The plane-wave state $\Delta \propto \exp(i\bm k \bm r_{||})$ can, in principle, carry a supercurrent in the bilayer plane. It is interesting to calculate this supercurrent. The corresponding expression for the supercurrent density takes the form
\begin{equation}
j^{(0)}(x)=\frac{\sigma(x)}{4\pi^2e}\bm k\int \limits_{-\infty}^\infty d \varepsilon  {\rm Im}\left\{ {f^{(0)}}^2(x)\right\} (\varphi_\uparrow+\varphi_\downarrow) 
\label{zero_current}
\enspace ,
\end{equation}
where $\sigma(x)=\sigma_{S(N)}$ in the S(N) layer, $f^{(0)}(x)$ is the solution of the linearized Usadel equation, expressed by Eqs.~(\ref{fS})-(\ref{fN}). The superscript $(0)$ means that it is calculated in the absence of the magnetic field. It is well-known \cite{fulde64} that for a homogeneous system the true ground state corresponds to zero current density. For our bilayer system this statement is valid for the total current, integrated over the bilayer width $\int \limits_{-d_N}^{d_S} dx j(x)=0 $. It can be shown by straightforward calculations that this is valid simultaneously with  $\partial T_c/\partial k^2=0$, that is at $k=k_{opt}$. Vanishing of the total current means that the supercurrent mainly flows in the opposite directions in the N and S regions of the bilayer. This results in the appearance of the magnetic flux, which can be a hallmark of $\exp(i\bm k \bm r_{||})$ in the bilayer. This flux is plotted in Fig.~\ref{flux} vs $eV$. The spatial profile of the corresponding magnetic field is shown in the insert to Fig.~\ref{flux}. However, for a state $\propto \cos(i\bm k \bm r_{||})$ the supercurrent density $j(x)=0$ locally for a given $x$. Consequently, this state is not accompanied by the non-zero supercurrents and cannot be detected by the corresponding magnetic flux.

Now we turn to the calculation of the Meissner response of the bilayer in the limit of a weak magnetic field $\bm H$, applied in the plane of the bilayer. As it was shown in Ref.~\onlinecite{mironov12}, the transition of S/F hybrid structures to the in-plane FFLO-state is accompanied by vanishing of the Meissner effect. It is connected to the fact that the Meissner response of a S/F bilayer in the homogeneous state can become paramagnetic and such a structure is unstable with respect to the formation of the FFLO-state \cite{mironov12}. The homogeneous state of our nonmagnetic S/N bilayer never exhibits the paramagnetic Meissner response. So, vanishing of the Meissner response cannot be a hallmark of the in-plane FFLO-state in our system. However, we have found another features of the linear response, which are typical for a in-plane FFLO-state in heterostructures. 

We choose the vector potential $\bm A=(0, H_z x, -H_y x)$ to be parallel to the $yz$-plane. For the considered FFLO-state $\propto \exp(i\bm k \bm r_\parallel)$ the linear in magnetic field contribution to the electric current density takes the form
\begin{eqnarray}
j^{(1)}(x)=\frac{\sigma(x)}{2\pi^2e}\int \limits_{-\infty}^\infty d \varepsilon \left[ \bm k {\rm Im}\left\{ f^{(1)}(x)f^{(0)}(x) \right\} -\right. \nonumber \\ 
\left.\frac{e}{c}(\bm A-\frac{c}{2e}\delta \bm k^{(1)}) {\rm Im}\left\{ f^{(0)2}(x) \right\} \right](\varphi_\uparrow+\varphi_\downarrow) 
\enspace ,
\label{meissner}
\end{eqnarray}
where the vector potential is taken in the gauge invariant form $\bm A-\frac{c}{2e}\delta \bm k^{(1)}$. $f^{(1)}(x)$ is the linear correction to the anomalous Green's function. It is worth noting that in the homogeneous state $f^{(1)}(x)$ is zero in the gauge ${\rm div} \bm A=0$. This is because ${\rm div} \bm A$ is the only possible first order scalar function of $\bm A$. In the FFLO-state $f^{(1)}(x) \propto \bm k \bm A'_x$. Therefore, in general, the linear response of the heterostructure in the FFLO-state can be anisotropic with respect to the direction of the applied magnetic field \cite{note0}. 

The full expression for $f^{(1)}(x)$ takes the form
\begin{equation}
f^{(1)}(x)=\frac{\Delta^{(1)}}{\Delta^{(0)}} f^{(0)}(x)+F^{(1)}_S
\label{delta1}
\enspace ,
\end{equation}
where
\begin{eqnarray}
 F^{(1)}_S = \frac{(2e/c)(i D_S/d_S)}{E} \left[ f^{(0)}_S\int \limits_0^{d_S} \bm k (\bm A(x)-\delta \bm k^{(1)})d x +\right. \nonumber \\
\left. \frac{\frac{g_{NS}}{\sigma_S}\int \limits_{-d_N}^{0} \bm k (\bm A(x)-\delta \bm k^{(1)})f^{(0)}_N(x)\frac{\cosh[\lambda(x+d_N)]}{\cosh[\lambda d_N]}dx}{\lambda \tanh [\lambda d_N]+g_{NS}/\sigma_N}  \right]~~~ \label{FS1}
\end{eqnarray}
and
\begin{equation}
E=\varepsilon + iD_S k^2/2 + \frac{ig_{NS}D_S\lambda \tanh [\lambda d_N]}{2\sigma_S d_s(\lambda \tanh [\lambda d_N]+g_{NS}/\sigma_N)}
\label{E}
\enspace . 
\end{equation}

The linear in magnetic field correction $\Delta^{(1)}$ to the superconducting order parameter can be obtained from the self-consistency equation Eq.~(\ref{Tc}). The expression for $\Delta^{(1)}$ takes the form  
\begin{equation}
\Delta^{(1)} = \frac{\int \limits_{-\omega_D}^{\omega_D} \frac{d\varepsilon}{4\pi} \Lambda {\rm Im} \left[ F^{(1)}_S \right] (\varphi_\uparrow+\varphi_\downarrow)}{1-\int \limits_{-\omega_D}^{\omega_D} \frac{d\varepsilon}{4} \Lambda {\rm Re} \left[ 1/E\right] (\varphi_\uparrow+\varphi_\downarrow)}
\label{S1}
\enspace . 
\end{equation}
The denominator of Eq.~(\ref{S1}) vanishes at $T=T_c$ because it is just the equation for calculating $T_c$ at zero applied field. Consequently, for temperatures close to $T_c$ the linear correction $\Delta^{(1)} \propto \Delta^{(0)}/(T_c-T)$. Therefore, the main contribution to $f^{(1)}(x)$ is given by the first term $f^{(1)}_{\Delta}(x)\propto \Delta^{(1)}$ in Eq.~(\ref{delta1}). 
In the state $\propto \cos (\bm k \bm r_{\parallel})$ the leading contribution to the Meissner current takes the same form (it is only two times larger). Certainly, this behavior violates extremely close to $T_c$, where $\Delta^{(1)}$ becomes of the order of $\Delta^{(0)}$ and our linear approximation fails.

Therefore, as it follows from Eq.~(\ref{meissner}), the Meissner response of the S/N bilayer system in the FFLO-state would exhibit non-trivial temperature dependence. While in the homogeneous state the Meissner current $j^{(1)}(T) \propto \Delta^2 \propto (T_c-T) $ if the temperature is near $T_c$, in the FFLO-state the leading contribution to $j^{(1)}(T) \propto \Delta^{(1)}\Delta^{(0)} $ and does not depend on temperature. In fact, this means that there are two possibilities: (i) the temperature dependence of the Meissner response near $T_c$ in the FFLO-state will be indeed non-trivial or (ii) $T_c$ itself is shifted by the magnetic field in the FFLO-state in the linear approximation, but the temperature dependence of the Meissner response can be of standard type. At the same time $T_c$ of the homogeneous system does not depend on the applied magnetic field in the linear approximation. Which of the possibilities is realized in the particular system depends on what type of the FFLO-state is more stable in the system (plane wave, stationary wave, etc.). In any case near $T_c$ the behavior of the linear response of the system on the applied magnetic field in the FFLO-state strongly differs from the behavior of the same system in the homogeneous state.  

Anisotropy of $\Delta^{(1)}$ with respect to the mutual direction of the applied magnetic field and the modulation vector $\bm k$ is also clearly seen from Eqs.~(\ref{S1}) and (\ref{FS1}). This, in turn, leads to the corresponding anisotropy of the Meissner response.  
 
In conclusion, we have shown that the in-plane FFLO-state can be stabilized in the S/N bilayer under non-equilibrium quasiparticle distribution for temperatures close to $T_c$. Its existence does not require any Zeeman interaction in the system. In general, this FFLO-state can be of different types: plane wave, stationary wave and, even, 2D-structures are possible. The plane wave state is accompanied by the internal magnetic flux. For all types of the FFLO-state near $T_c$  temperature dependence of the linear response of the system on the applied magnetic field should be strongly nontrivial.

{\it Acknowledgments.} The authors are grateful to S. Mironov for useful discussions. The support by RFBR Grant No. 
12-02-00723-a is acknowledged.



\begin{thebibliography}{99}
%
\bibitem{bianchi03}
A. Bianchi, R. Movshovich, C. Capan, P.G. Pagliuso, and J. L. Sarrao,  Phys. Rev. Lett. {\bf 91}, 187004 (2003).
%
\bibitem{capan04}
C. Capan, A. Bianchi, R. Movshovich, A. D. Christianson, A. Malinowski, M. F. Hundley, A. Lacerda, P.G. Pagliuso, and J. L. Sarrao, Phys. Rev. B {\bf 70}, 134513 (2004).
%
\bibitem{uji01}
S. Uji, H. Shinagawa, T. Terashima, T. Yakabe, Y. Terai, M. Tokumoto, A. Kobayashi, H. Tanaka and H. Kobayashi, Nature (London) {\bf 410}, 908 (2001).
%
\bibitem{buzdin05}
A.~I.~Buzdin, Rev. Mod. Phys. {\bf 77},  935 (2005).
%
\bibitem{bergeret05}
F.S. Bergeret, A.F. Volkov, and K.B. Efetov, Rev. Mod. Phys. {\bf 77}, 1321 (2005).
%
\bibitem{larkin64}
A.I. Larkin and Yu.N. Ovchinnikov, Sov. Phys. JETP {\bf 20}, 762 (1965) [Zh. Eksp. Teor. Fiz. {\bf 47}, 1136 (1964)].
%
\bibitem{fulde64}
P. Fulde and R.A. Ferrel, Phys.Rev. {\bf 135}, A550 (1964).
%
\bibitem{sarma63}
G. Sarma, J. Phys. Chem. Solids {\bf 24}, 1029 (1963).
%
\bibitem{maki68}
K. Maki, Progr. Theoret. Phys. {\bf 39}, 897 (1968).
%
\bibitem{singleton00}
J. Singleton, J.A. Symington, M.-S. Nam, A. Ardavan, M. Kurmoo, and P. Day, J. Phys.: Condens. Matter {\bf 12}, L641 (2000).  
%
\bibitem{tanatar02}
M. A. Tanatar, T. Ishiguro, H. Tanaka, and H. Kobayashi,  Phys. Rev. B {\bf 66}, 134503 (2002). 
%
\bibitem{bianchi02}
A. Bianchi, R. Movshovich, N. Oeschler, P. Gegenwart, F. Steglich, J.D. Thompson, P.G. Pagliuso, and J.L. Sarrao, Phys. Rev. Lett. {\bf 89}, 137002 (2002). 
%
\bibitem{miclea06}
C.F. Miclea, M. Nicklas, D. Parker, K. Maki, J. L. Sarrao, J. D. Thompson, G. Sparn, and F. Steglich, Phys. Rev. Lett. {\bf 96}, 117001 (2006). 
%
\bibitem{uji06}
S. Uji, T. Terashima, M. Nishimura, Y. Takahide, T. Konoike, K. Enomoto, H. Cui, H. Kobayashi, A. Kobayashi, H. Tanaka, M. Tokumoto, E.S. Choi, T. Tokumoto, D. Graf, and J.S. Brooks, Phys. Rev. Lett. {\bf 97}, 157001 (2006). 
%
\bibitem{shinagawa07}
J. Shinagawa, Y. Kurosaki, F. Zhang, C. Parker, S.E. Brown, D. Jerome, K. Bechgaard, and J. B. Christensen, Phys. Rev. Lett. {\bf 98}, 147002 (2007).
%
\bibitem{lortz07}
R. Lortz, Y. Wang, A. Demuer, P.H.M. Bottger, B. Bergk, G. Zwicknagl, Y. Nakazawa, and J. Wosnitza, Phys. Rev. Lett. {\bf 99}, 187002 (2007). 
%
\bibitem{cho09}
K. Cho, B.E. Smith, W.A. Coniglio, L.E. Winter, and C.C. Agosta, J.A. Schlueter, Phys. Rev. B {\bf 79}, 220507(R) (2009). 
%
\bibitem{wright11}
J.A. Wright, E. Green, P. Kuhns, A. Reyes, J. Brooks, J. Schlueter, R. Kato, H. Yamamoto, M. Kobayashi, and S.E. Brown, Phys. Rev. Lett. {\bf 107}, 087002 (2011). 
%
\bibitem{bergk11}
B. Bergk, A. Demuer, I. Sheikin, Y. Wang, J. Wosnitza, Y. Nakazawa, and R. Lortz, Phys. Rev. B {\bf 83}, 064506 (2011). 
%
\bibitem{coniglio11}
W.. A. Coniglio, L.E. Winter, K. Cho, C.C. Agosta, B. Fravel and L.K. Montgomery, Phys. Rev. B {\bf 83}, 224507 (2011). 
%
\bibitem{tarantini11}
C. Tarantini, A. Gurevich, J. Jaroszynski, F. Balakirev, E. Bellingeri, I. Pallecchi, C. Ferdeghini, B. Shen, H.H. Wen, and D.C. Larbalestier, Phys. Rev. B {\bf 84}, 184522 (2011). 
%
\bibitem{gebre11}
T. Gebre, G. Li, J.B. Whalen, B.S. Conner, H.D. Zhou, G. Grissonnanche, M.K. Kostov, A. Gurevich, T. Siegrist, and L. Balicas, Phys. Rev. B {\bf 84}, 174517 (2011). 
%
\bibitem{agosta12}
C.C. Agosta, Jing Jin, W.A. Coniglio, B.E. Smith, K. Cho, I. Stroe, C. Martin, S.W. Tozer, T.P. Murphy, E.C. Palm, J.A. Schlueter, 
M. Kurmoo, Phys. Rev. B {\bf 85}, 214514 (2012). 
%
\bibitem{uji12}
S. Uji, K. Kodama, K. Sugii, T. Terashima, Y. Takahide, N. Kurita, S. Tsuchiya, M. Kimata, A. Kobayashi, B. Zhou, and H. Kobayashi, Phys. Rev. B {\bf 85}, 174530 (2012). 
%
\bibitem{mironov12}
S. Mironov, A. Mel'nikov, and A. Buzdin, Phys. Rev. Lett. {\bf 109}, 237002 (2012). 
%
\bibitem{sidorenko09}
A.S. Sidorenko, V.I. Zdravkov, J. Kehrle, R. Morari, G. Obermeier, S. Gsell, M. Schreck, C. Muller, M.Yu. Kupriyanov, V.V. Ryazanov, S. Horn, L.R. Tagirov, and R. Tidecks, JETP Lett. {\bf 90}, 139 (2009).
%
\bibitem{pothier97}
H. Pothier, S. Gueron, N.O. Birge, D. Esteve, and M.H. Devoret, Phys. Rev. Lett. {\bf 79}, 3490 (1997).
%
\bibitem{baselmans99}
J.J.A. Baselmans, A.F. Morpurgo, B.J. van Wees, and T.M. Klapwijk, Nature (London) {\bf 397}, 43 (1999).
%
\bibitem{doh06}
H. Doh, M. Song, and H.-Y. Kee, Phys. Rev. Lett. {\bf 97}, 257001 (2006).  
%
\bibitem{vorontsov09}
A.B. Vorontsov, Phys. Rev. Lett. {\bf 102}, 177001 (2009).  
%
\bibitem{volkov09}
A. Moor, A.F. Volkov and K.B. Efetov, Phys. Rev. B {\bf 80}, 054516 (2009).  
%
\bibitem{saint-james69}
D. Saint-James, G. Sarma and E.J. Thomas, Type II superconductivity (Pergamon, Oxford, 1969).
%
\bibitem{bobkov13}
A.M. Bobkov and I.V. Bobkova, arXiv:1309.2461.
%
\bibitem{usadel}
K.D. Usadel, Phys.Rev.Lett. {\bf 25}, 507 (1970).
%
\bibitem{kupriyanov88}
M.Yu. Kupriyanov and V.F. Lukichev, Sov. Phys. JETP {\bf 67}, 1163
(1988).
%
\bibitem{aronov76}
A.G. Aronov, JETP Lett. {\bf 24}, 32 (1976).
%
\bibitem{tsoi}
M.V. Tsoi and V.S. Tsoi, JETP Lett. {\bf 73}, 98 (2001).
%
\bibitem{note1}
The parameters are chosen to be close to the real experimental parameters for Nb/Cu bilayers.
%
\bibitem{note0}
It was already proposed in the literature to detect the magnetic field-driven FFLO-state by the anisotropy of $T_c$ in dependence on the magnetic field direction:  M.D.Croitoru, M.Houzet, A.I.Buzdin, Phys. Rev. Lett. {\bf 108}, 207005 (2012).      
%
\end{thebibliography}

\end{document}